\documentstyle[preprint,prd,aps,floats]{revtex}

\renewcommand{\to}{\rightarrow}
\newcommand{\as}{\alpha_s}
\newcommand{\Ei}{{\rm Ei}}
\newcommand{\kv}{{\bf k}}
\newcommand{\sv}{{\bf S}}
\newcommand{\pv}{{\bf p}}
\newcommand{\rv}{{\bf r}}
\begin{document}
\thispagestyle{empty}
\draft
\title{Heavy Quarkonium Potential Model \\and the ${}^1P_1$ State of
Charmonium}
\author{Suraj N. Gupta and James M. Johnson}
\address{ Department of Physics, Wayne State University, Detroit,
	Michigan 48202}
\author{Wayne W. Repko}
\address{Department of Physics and Astronomy, Michigan State University,
	East Lansing, Michigan 48824}
\author{Casimir J. Suchyta III}
\address{Cray Research, Inc., 655F Lone Oak Drive, Eagan, Minnesota 55121}
\maketitle
\begin{abstract}
	A theoretical explanation of the observed splittings among the
P~states of charmonium is given with the use of a nonsingular potential
model for heavy quarkonia.  We also show that the recently observed mass
difference between the center of gravity of the ${}^3P_J$ states and the
${}^1P_1$ state of $c\bar{c}$ does not provide a direct test of the
color hyperfine interaction in heavy quarkonia.  Our theoretical value
for the mass of the ${}^1P_1$ state is in agreement with the experimental
result,
and its E1 transition width is 341.8~keV.  The mass of the $\eta_c'$ state
is predicted to be 3622.3~MeV.
\end{abstract}
\narrowtext
\raggedright
\setlength{\parindent}{24pt}
\section{INTRODUCTION}

	A quantum-chromodynamic potential model was proposed by us\cite{GRR}
in 1982, which not only yielded results for the $c\bar{c}$ and $b\bar{b}$
energy levels and their spin splittings in good agreement with the existing
experimental data but its predictions were also confirmed by
later experiments at the Cornell Electron Storage Ring\cite{CESR}.
An essential feature of our model was the inclusion of the
one-loop radiative corrections to the quark-antiquark potential, which
had been derived by us in an earlier investigation\cite{GR}. Subsequently,
the model was improved by using relativisitic kinematics\cite{GRR2}
and a nonsingular form of the quarkonium potential\cite{GRS}.
As shown by us, in addition to the energy levels of $c\bar{c}$ and $b\bar{b}$,
our model also yields results in good agreement with the experimental data
for the leptonic and E1 transition widths.  It was further shown by
Zhang, Sebastian, and Grotch\cite{zhang} that the M1 transition
widths for $c\bar{c}$ and $b\bar{b}$ obtained from our model
are in better agreement with the experimental data than those predicted
using other potential models.

	Recently the mass of the ${}^1P_1$ state of charmonium has
been determined by the E760 collaboration\cite{E760} in $p\bar{p}$
annihilations at Fermilab, and the splitting between the center of
gravity of the ${}^3P_J$ states and the ${}^1P_1$ state, denoted as
$\Delta M_P$, is found to be approximately $-0.9$~MeV.  This experimental
result has created much interest since it provides a new test for the
potential models for heavy quarkonia.

	If the spin-dependent forces in the quarkonium potential could be
treated perturbatively, the $\Delta M_P$ splitting would arise solely
from the spin-spin (color hyperfine) interaction.  However, the spin-dependent
forces are known to be quite large and, as observed by Lichtenberg and Potting
\cite{lichten}, the contributions of the spin-orbit and tensor interactions to
$\Delta M_P$ cannot be ignored in a nonperturbative treatment.  We shall
analyze this complex situation with the use of our model which
avoids the use of an illegitimate perturbative treatment, and provide an
explanation for the observed splittings of the charmonium P~states.

	Several authors\cite{halzen,chen,grotch} have recently shown that
a theoretical value for $\Delta M_P$ in close agreement with the
experimental value can be readily obtained from the spin-spin
interaction terms in the quarkonium potential.  However,
since they have employed an illegitimate perturbative treatment, the
significance of this simple interpretation remains an open question.

	Only a quarkonium model which is in good overall agreement with
the experimental data can be taken seriously. Our model for heavy quarkonia
satisfies this requirement.

\section{$\symbol{'143}\bar{\symbol{'143}}$ SPECTRUM}

	Our model is based on the Hamiltonian
\begin{equation}
H=H_0+V_p+V_c , \label{hamiltonian}
\end{equation}
where
\begin{equation}
H_0=2(m^2+{\bf p}^2)^{1/2}
\end{equation}
is the relativistic kinetic energy term, and $V_p$ and $V_c$ are nonsingular
quasistatic perturbative and confining potentials, which are
given in the Appendix.
We found a trial wave function introduced by Jacobs, Olsson, and Suchyta
\cite{jacobs} particularly suitable for obtaining the quarkonium energy levels
and wave functions.

	Our results for the energy-level splittings as well as the individual
energy levels of $c\bar{c}$ are given in Tables~\ref{splittings}
and~\ref{levels}.
For experimental data we have generally relied on the Particle Data Group
\cite{PDG}, but for the $\eta_c$ state we have used the new result announced by
the E760 collaboration \cite{appel}.  The two sets
of theoretical results in these tables correspond to the scalar-exchange and
the scalar-vector-exchange forms of the confining
potential, given by
\begin{mathletters}
\begin{equation}
V_c=V_S ,
\end{equation}
and
\begin{equation}
V_c=(1-B)V_S+BV_V ,
\label{vconfine}
\end{equation}
\end{mathletters}
respectively.  The results obtained with the scalar-exchange confining
potential are
unsatisfactory, while the scalar-vector-exchange results
are in surprisingly close agreement with the
experimental data, including the observed mass of the ${}^1P_1$ state
and the $\Delta M_P$ splitting.  The scalar-vector mixing parameter $B$
is found to be about $\frac{1}{4}$.

	In Table~\ref{mp}, we display the contributions to $\Delta M_P$ from
the various types of terms in the Hamiltonian (\ref{hamiltonian}) with
the confining potential (\ref{vconfine}).
The table shows comparable contributions to $\Delta M_P$
from several sources, which brings out the complexity of this splitting when
spin-dependent potential terms are included in the unperturbed Hamiltonian.
The $\Delta M_P$ splitting, therefore, does not provide a direct test of the
spin-spin interaction in heavy quarkonia.

	In Tables~IV and~V, we give the results for the leptonic and E1 transition
widths corresponding to the scalar-vector-exchange confining potential by using
the formulae
\begin{equation}
\Gamma_{ee}({}^3S_1 \to e^+e^-) = \frac{16\pi\alpha^2
e_Q^2}{M^2(Q\overline{Q})}
	\left| \Psi(0)\right|^2 \left( 1-\frac{16\alpha_s}{3\pi}\right),
\end{equation}
and
\begin{eqnarray}
\Gamma_{E1}({}^3S_1 \to {}^3P_J) &=& \frac{4}{9}\frac{2J+1}{3}\alpha e_Q^2
	k_J^3 |r_{fi}|^2,\nonumber \\
\Gamma_{E1}({}^3P_J \to {}^3S_1) &=&\frac{4}{9}\alpha e_Q^2
	k_J^3 |r_{fi}|^2,\\
\Gamma_{E1}({}^1P_1 \to {}^1S_0) &=&\frac{4}{9}\alpha e_Q^2
	k_J^3 |r_{fi}|^2.\nonumber
\end{eqnarray}
The photon energies for the E1 transition widths have been
obtained from the energy difference of the initial and the
final $c\bar{c}$ states by taking into account the recoil correction.
Our results are
in good agreement with the available experimental data \cite{PDG}, and
our prediction for the E1 transition width of $1{}^1P_1\to 1{}^1S_0$ is
341.8 keV.

\section{CONCLUSION}
	We conclude with explanatory remarks concerning some features of our
quarkonium potential.

\subsection{Renormalization scheme}

	We have used the Gupta-Radford (GR) renormalization scheme \cite{GR2}
for the one-loop radiative corrections to the quarkonium potential
rather than the modified minimal-subtraction ($\overline{\rm MS}$)
scheme.  The GR scheme is a simplified momentum-space subtraction
scheme, and the parameter $\mu$ can be interpreted as representing the
momentum scale of the physical process.
This scheme also has the desirable feature that
it satisfies the decoupling theorem\cite{appelquist}.  On the other
hand, in the $\overline{\rm MS}$ scheme $\mu$ appears as a mathematical
parameter, and in this scheme decoupling-theorem-violating terms are
simply ignored.

	The one-loop radiative corrections in the GR scheme can be converted
into those in the $\overline{\rm MS}$ scheme by means of the relation
\cite{GR2}
\begin{equation}
\alpha_s=\bar{\alpha_s}\left[ 1+\frac{\bar{\alpha_s}}{4\pi}\left(
	\frac{49}{3}-\frac{10}{9}n_l+\frac{2}{3}\sum_{n_h}
	\ln\frac{m^2}{\mu^2} \right)\right]\, , \label{renorm}
\end{equation}
where $\bar{\alpha_s}$ refers to the $\overline{\rm MS}$ scheme, and $n_l$
and $n_h$ are the numbers of light and heavy quark flavors.  If we drop
the decoupling-theorem-violating terms that appear in the $\overline{\rm MS}$
scheme, we can put $n_l=n_f$ and $n_h=0$, and (\ref{renorm}) reduces to
\begin{equation}
\alpha_s=\bar{\alpha_s}\left[ 1+\frac{\bar{\alpha_s}}{4\pi}\left(
	\frac{49}{3}-\frac{10}{9}n_f \right)\right]\, .
\end{equation}

\subsection{Quasistatic potential}

	In an earlier investigation \cite{GRR2}, we arrived at the surprising
conclusion that while the quasistatic form of the quarkonium potential yields
results in good agreement with the experimental data, this is not the case
for the momentum-dependent form.  This conclusion has also been confirmed by
the
recent investigations of Gara {\it et al.} \cite{gara} and Lucha {\it et al.}
\cite{lucha}.

	It appears to us that the success of the quasistatic potential is related
to the phenomenon of quark confinement.  Since a rigorous treatment of
quark confinement does not exist at this time, we shall only offer a
plausible argument. It was argued earlier \cite{GR3} with the use
of a renormalization-group-improved quantum-chromodynamic treatment
that quark confinement can be understood as a consequence of the fact
that quarks and antiquarks are unable to exchange low-momentum gluons.
Moreover, since for the quark-antiquark scattering in the center-of-mass
system
\begin{equation}
{\bf p}^2=\frac{1}{4}{\bf k}^2+\frac{1}{4}{\bf s}^2,
\end{equation}
where
\begin{equation}
{\bf k}={\bf p'}-{\bf p},\qquad {\bf s}={\bf p'}+{\bf p},
\end{equation}
it follows that if ${\bf k}^2$ is allowed to take only large values,
${\bf s}^2$ can be treated as small.  This may be regarded as a justification
for the quasistatic approximation in which terms of second and higher
orders in ${\bf s}$ are ignored.

	Our quarkonium perturbative and confining potentials are not
only quasistatic but also nonsingular.  In the momentum space, these
potentials are obtained by first expanding in powers of
$\pv^2/(m^2+\pv^2)$, and then approximating $\pv^2$ as $\frac{1}{4}\kv^2$.
The perturbative potential in powers of $\pv^2/(m^2+\pv^2)$ includes, among
others, terms of the form
\begin{equation}
f(\pv^2)=\frac{a+b\;\sv_1\cdot\sv_2}{m^2+\pv^2}\ ,
\label{zero}
\end{equation}
which becomes in the quasistatic approximation
\begin{equation}
f(\kv^2)=\frac{a+b\;\sv_1\cdot\sv_2}{m^2+\frac{1}{4}\kv^2}\ .
\label{quasizero}
\end{equation}
It has been observed by Grotch, Sebastian, and Zhang \cite{grotch} that while
the
contribution of $f(\pv^2)$ vanishes for the P states due to the
vanishing of the wave function at the origin, $f(\kv^2)$ yields a small
but nonvanishing contribution for these states.  Consequently, for P and higher
angular-momentum states it would be more accurate to drop
terms of the form (\ref{zero}) than to convert them into the approximate form
(\ref{quasizero}).  We agree with the observation of Grotch {\it et al.}
Accordingly, in the treatment of states with $l\neq 0$ we shall
drop terms of the form (\ref{quasizero}) in the momentum-space potentials
and the corresponding terms of the form
\begin{equation}
f(\rv)=\frac{a+b\;\sv_1\cdot\sv_2}{\pi r}e^{-2mr}
\label{coordzero}
\end{equation}
in the coordinate-space potentials.

\subsection{Confining potential}

	In our theoretical treatment, our aim has been to avoid phenomenology
except in the choice of the long-range confining potential, which
cannot be derived sufficiently accurately by any known theoretical
technique.  It is indeed remarkable that the results obtained from our
field-theoretical perturbative potential supplemented with a phenomenological
confining potential are in
excellent over-all agreement with the experimental data including the
$\Delta M_P$ splitting. It should be noted that we have neglected
effect of coupling of the energy levels to virtual decay
channels and possibly other small effects.
Such effects presumably have also been taken into account in
our phenomenological confining potential.

\acknowledgments
	This work was supported in part by the U.S. Department of Energy
under Grant No. DE-FG02-85ER40209 and the National Science Foundation
under Grant No. PHY-93-07980. W.~W.~R. would like to acknowledge conversations
with R.~Lewis and G.~A.~Smith regarding the results of the E760
collaboration.

\cleardoublepage
\widetext
\appendix
\section*{NONSINGULAR QUARKONIUM POTENTIALS}

	The nonsingular quarkonium potentials can be obtained \cite{SNG} by
appropriate
modifications of the singular potentials in the momentum space,
and transforming them to the coordinate space.  The nonsingular potentials
obtained by this procedure are given below.  Some unwanted terms for states
with
$l\neq 0$ have been dropped as explained in Sec.~III.

\subsection{Perturbative quantum-chromodynamic potential}
	The perturbative potential $V_p$ consists of the direct potential $V_p'$
and the annihilation potential $V_p''$, and in the momentum space
\begin{equation}
V_p({\bf k})=V_p'({\bf k})+V_p''({\bf k}),
\end{equation}
where
\begin{eqnarray}
V_p'({\bf k})&=& -\frac{16\pi\as}{3\kv^2}\left[1-\frac{3\as}{2\pi}
	-\frac{\as}{12\pi}(33-2n_f)\ln\left(\frac{\kv^2}{\mu^2}\right)
		\right]\nonumber\\
& & \hspace*{0.5in}+\frac{16\pi\as}{3(\kv^2+4m^2)}\left[
	\delta_{l0}\left(1-\frac{3\as}{2\pi}\right)
	-\frac{\as}{12\pi}(33-2n_f)\ln\left(\frac{\kv^2}{\mu^2}\right)
	-\frac{7\pi\as}{3}\;\frac{m}{|\kv|}\right]\nonumber\\[8pt]
& & \hspace*{0.in}+\frac{128\pi\as}{9}\;\frac{\sv_1\!\cdot\!\sv_2}{\kv^2+4m^2}
	\left[\delta_{l0}\left(1-\frac{35\as}{12\pi}\right)
	-\frac{\as}{12\pi}(33-2n_f)\ln\left(\frac{\kv^2}{\mu^2}\right)
	+\frac{21\as}{8\pi}\ln\left(\frac{\kv^2}{m^2}\right)
	\right]\nonumber\\[8pt]
& & \hspace*{0.in}-32\pi\as\frac{i\sv\cdot(\kv\times\pv)}{\kv^2(\kv^2+4m^2)}
	\left[1-\frac{11\as}{18\pi}
	-\frac{\as}{12\pi}(33-2n_f)\ln\left(\frac{\kv^2}{\mu^2}\right)
	+\frac{\as}{\pi}\ln\left(\frac{\kv^2}{m^2}\right)
		\right]\nonumber\\[8pt]
& & \hspace*{0.in}-\frac{64\pi\as}{3}\;
	\frac{\sv_1\!\cdot\!\kv\ \sv_2\!\cdot\!\kv
	-\frac{1}{3}\kv^2\sv_1\!\cdot\!\sv_{2} }{\kv^2(\kv^2+4m^2)}
	\left[1+\frac{4\as}{3\pi}
	-\frac{\as}{12\pi}(33-2n_f)\ln\left(\frac{\kv^2}{\mu^2}
		\right)\right.\nonumber\\
& & \hspace*{2.5in}\left.+\frac{3\as}{2\pi}\ln\left(\frac{\kv^2}{m^2}
	\right)\right]\ ,\\[12pt]
V_p''({\bf k})&=& \delta_{l0}\frac{32\as^2}{3(\kv^2+4m^2)}
	\left(1-\ln 2\right)\left(\sv_1\cdot\sv_2-\frac{1}{4}\right).
\end{eqnarray}

	In the coordinate space, the potential takes the form
\begin{equation}
V_p({\bf r})=V_p'({\bf r})+V_p''({\bf r}),
\end{equation}
where
\begin{eqnarray}
V_p'({\bf r})&=& -\frac{4\as}{3r}\left\{1-\frac{3\as}{2\pi}+\frac{\as}{6\pi}
	(33-2n_f)\left[\ln(\mu r)+\gamma_E\right]\right\}\nonumber\\[8pt]
& & +\frac{4\as}{3r}\left\{\delta_{l0}\left(1-\frac{3\as}{2\pi}\right)e^{-2mr}
	+\frac{\as}{6\pi}(33-2n_f)\left[\ln(\mu r)e^{-2mr}
		+E_+(2mr)\right]\right.\nonumber \\
& &\hspace*{0.75in}-\left. \frac{7\as}{3}\left[\ln(2mr)e^{-2mr}-E_-(2mr)
		\right]
		\right\}\nonumber\\[8pt]
& & +\frac{32\as}{9r}\sv_1\cdot\sv_2\left\{ \delta_{l0}\left(1-
	\frac{35\as}{12\pi}\right)e^{-2mr}+\frac{\as}{6\pi}(33-2n_f)
	\left[\ln(\mu r)e^{-2mr}+E_+(2mr)\right]\right.\nonumber\\
& & \hspace*{0.75in}\left. -\frac{21\as}{4\pi}\left[
	\ln(mr)e^{-2mr}+E_+(2mr)\right]\right\} \nonumber\\[8pt]
& & +\frac{8\as}{r}{\bf L}\cdot\sv\left\{\left(1-\frac{11\as}{18\pi}\right)
	f_1(2mr)+\frac{\as}{6\pi}(33-2n_f)\left[f_1(2mr)\ln(\mu r)
	+g_1(2mr)\right]\right.\nonumber\\
& & \hspace*{0.75in}\left. -\frac{2\as}{\pi}\left[f_1(2mr)\ln(mr)+g_1(2mr)
		\right]\right\}\nonumber\\[8pt]
& & +\frac{4\as}{3r}S_T\left\{\left(1+\frac{4\as}{3\pi}\right)f_2(2mr)
	+\frac{\as}{6\pi}(33-2n_f)\left[f_2(2mr)\ln(\mu r)+g_2(2mr)
		\right]\right.\nonumber\\
& & \hspace*{0.75in}\left. -\frac{3\as}{\pi}\left[f_2(2mr)\ln(mr)+g_2(2mr)
		\right]\right\},\\[12pt]
V_p''({\bf r})&=& \delta_{l0}\frac{8\as^2e^{-2mr}}{3\pi r}
	\left(1-\ln 2\right)\left(\sv_1\cdot\sv_2-\frac{1}{4}\right) .
\end{eqnarray}
\narrowtext

	Note that the tensor operator is defined as
\begin{equation}
S_T=3\ \mbox{\boldmath$\sigma$}_1\cdot{\bf \hat{r}}\
	\mbox{\boldmath$\sigma$}_2\cdot{\bf \hat{r}}
	-\mbox{\boldmath$\sigma$}_1\cdot\mbox{\boldmath$\sigma$}_2,
\end{equation}
the functions $E_\pm$ are expressible in terms of the
exponential-integral function $\Ei$ as
\begin{equation}
E_\pm(x)= \frac{1}{2}\left[e^x \Ei(-x)\pm e^{-x}\Ei(x)\right]
	\mp e^{-x}\ln x,
\end{equation}
and
\begin{eqnarray}
f_1&=& \frac{1-(1+x)e^{-x}}{x^2},\nonumber\\[2pt]
f_2&=& \frac{1-\left(1+x+\frac{1}{3}x^2\right)e^{-x}}{x^2},\\[2pt]
g_1&=& \frac{\gamma_E-\left[E_+(x)-xE_-(x)\right]}{x^2},\nonumber\\[2pt]
g_2&=& \frac{\gamma_E-\left[\left(1+\frac{1}{3}x^2\right)E_+(x)
	-xE_-(x)\right]}{x^2}.\nonumber
\end{eqnarray}

\subsection{Phenomenological confining potential}

	The scalar-exchange and the vector-exchange
confining potentials in the momentum space are
\begin{equation}
V_S({\bf k})= -8\pi A\left[\frac{1}{\kv^4}-\frac{2i \sv\cdot(\kv\times\pv)}{
	\kv^4(\kv^2+4m^2)}\right] ,
\end{equation}
and
\begin{equation}
V_V({\bf k})=  -8\pi A\left[\frac{1}{\kv^4}-\frac{1+\frac{8}{3}\sv_1
	\cdot\sv_2}{\kv^2(\kv^2+4m^2)}
	+\frac{6i\sv\cdot(\kv\times\pv)}{\kv^4(\kv^2+4m^2)}
	+4\frac{\sv_1\cdot\kv\;\sv_2\cdot\kv-\frac{1}{3}\kv^2
	\sv_1\cdot\sv_2}{\kv^4(\kv^2+4m^2)}\right] .
\end{equation}

	The coordinate-space potentials are given by
\begin{equation}
V_S({\bf r})= Ar-\frac{A}{2m^2r}{\bf L}\cdot\sv\left[1-2f_1(2mr)\right] ,
\end{equation}
and
\begin{eqnarray}
V_V({\bf r})&=& Ar+\frac{A}{2m^2r}\left(1+\frac{8}{3}\sv_1\cdot\sv_2\right)
	\left(1-e^{-2mr}\right)+\frac{3A}{2m^2r}{\bf L}\cdot\sv
	\left[1-2f_1(2mr)\right]\nonumber\\
& & \hspace{0.5in}+\frac{A}{12m^2r}S_T\left[1-6f_2(2mr)\right] .
\end{eqnarray}

It is understood that the confining potential also contains an additive
phenomenological constant $C$.

\cleardoublepage

\clearpage

\mediumtext
\begin{table}
\caption{\label{splittings} $c\bar{c}$ energy level splittings in MeV.
Theoretical splittings and parameters correspond to the scalar-exchange
and the scalar-vector-exchange forms of the
confining potential.  The experimental value of the $\psi'-\eta_C'$
splitting is not used for the determination of the $c\bar{c}$
parameters because of the uncertainty regarding the $\eta_c'$ mass. }
\bigskip
\centerline{ \parbox{5.0in}{
\begin{tabular}{lddr@{$\pm$}l}
\hspace*{18pt}&Scalar&Scalar-vector&\multicolumn{2}{c}{Expt.}\\
\tableline
$\psi'-J/\psi$& 	587.7&	588.9&	$589.07$&$0.13$ \\
$J/\psi-\eta_c$&	105.1& 	109.0&	$109.03$&$3.1$ \\
$\psi'-\eta_c'$&	60.5&	63.5&	\multicolumn{2}{c}{} \\
$\chi_{\rm cog}-J/\psi$&430.5&	428.6&	$428.35$&$1$ \\
$\chi_{c2}-\chi_{c1}$&28.6&	44.6&	$45.64$&$0.18$ \\
$\chi_{c1}-\chi_{c0}$&80.0&	95.8&	$95.43$&$1$ \\
$\chi_{\rm cog}-h_c$&$-$5.8&$-$0.9&	$-0.93$&$0.19\pm0.2$\\
$m_c$ (GeV)& 	2.375&	2.208& \multicolumn{2}{c}{}\\
$\mu$ (GeV)&	3.329&	2.580& \multicolumn{2}{c}{}\\
$\alpha_s$&	0.295& 	0.313& \multicolumn{2}{c}{}\\
$A$ (GeV$^2$)&	0.183& 	0.181& \multicolumn{2}{c}{}\\
$B$&		&	0.245& \multicolumn{2}{c}{}
\end{tabular} } }
\end{table}

\begin{table}
\caption{\label{levels} $c\bar{c}$ energy levels in MeV corresponding to the
scalar-exchange and the scalar-vector-exchange forms of the
confining potential. }
\bigskip
\centerline{ \parbox{5.0in}{
\begin{tabular}{lccr@{$\pm$}l}
\hspace*{1pt}&Scalar&Scalar-vector&\multicolumn{2}{c}{Expt.}\\
\tableline
$1{}^3S_1\, (J/\psi)$&	3096.9&	3096.9&		$3096.93$&$0.09$ \\
$1{}^1S_0\, (\eta_c)$&	2991.8&	2987.9&		$2987.9$&$3.1$ \\
$2{}^3S_1\, (\psi')$&	3684.6&	3685.8&		$3686.0$&$0.1$ \\
$2{}^1S_0\, (\eta_c')$&	3624.1&	3622.3&		\multicolumn{2}{c}{} \\
$1{}^3P_2\, (\chi_{c2})$& 3549.0& 3556.0&	$3556.17$&$0.13$ \\
$1{}^3P_1\, (\chi_{c1})$& 3520.4& 3511.3&	$3510.53$&$0.12$ \\
$1{}^3P_0\, (\chi_{c0})$& 3440.4& 3415.5&	$3415.1$&$1$ \\
$1{}^1P_1\, (h_c)$&	3533.2&	3526.3&		$3526.2$&$0.15\pm0.2$
\end{tabular} } }
\end{table}
\vspace{1in}

\narrowtext
\begin{table}
\caption{\label{mp} Contributions to the $\chi_{\rm cog}-{}^1P_1$
splitting in MeV from the
various types of terms in the $c\bar{c}$ Hamiltonian.  The spin-independent,
spin-spin, spin-orbit, and tensor potential terms are denoted as
$V_{SI}$, $V_{SS}$, $V_{LS}$, and $V_T$, respectively. }
\bigskip
\centerline{ \parbox{2.5in}{
\begin{tabular}{cd}
 Hamiltonian term&$\chi_{\rm cog}-{}^1P_1$\\
\tableline
$H_0$&		19.4\\
$V_{SI}$&	$-$9.6\\
$V_{SS}$&	5.2\\
$V_{LS}$&	$-$13.7\\
$V_T$&		$-$2.2\\
\hline
Total&		$-$0.9
\end{tabular} } }
\end{table}
\clearpage

\begin{table}
\caption{\label{leptonic} $c\bar{c}$ leptonic widths in keV. }
\bigskip
\centerline{ \parbox{3in}{
\begin{tabular}{cdr@{$\pm$}l}
State&$\Gamma_{ee}$ (theory)&\multicolumn{2}{c}{$\Gamma_{ee}$ (expt.)}\\
\tableline
$1{}^3S_1$&	6.68&	$5.36$&$0.29$ \\
$2{}^3S_1$&	3.25&	$2.14$&$0.21$
\end{tabular} } }
\end{table}

\vspace{1in}
\mediumtext
\begin{table}
\caption{\label{E1} $E1$ transition widths for $c\bar{c}$ in keV.
The matrix elements $|r_{fi}|$ for these transitions are given in GeV$^{-1}$. }
\bigskip
\centerline{ \parbox{4in}{
\begin{tabular}{ccddr@{$\pm$}l}
Transition&$J$&$|r_{fi}|$&$\Gamma_{E1}$ (theory)&
	\multicolumn{2}{c}{$\Gamma_{E1}$ (expt.)}\\
\tableline
$2{}^3S_1\to 1{}^3P_J$&	2&	2.19&	24.2&	$21.7$&$3.3$ \\
		&	1&	1.96&	28.1&	$24.2$&$3.6$ \\
		&	0&	1.47&	18.3&	$25.9$&$3.9$ \\
$1{}^3P_J\to 1{}^3S_1$&	2&	1.60&	293.5&	$270.0$&$33$ \\
		&	1&	1.62&	225.2&	$240.0$&$41$ \\
		&	0&	1.62&	105.2&	$92.4$&$42$ \\
$1{}^1P_1\to 1{}^1S_0$&	&	1.39&	341.8& 	\multicolumn{2}{c}{}
\end{tabular} } }
\end{table}

\end{document}